\documentstyle[sprocl,psfig]{article}
\newcommand{\s}{\rm}

\newcommand{\ra}{\rightarrow}
\newcommand{\mn}{\mu \nu}

\newcommand{\be}{\begin{equation}}
\newcommand{\ee}{\end{equation}}

\newcommand{\bea}{\begin{eqnarray}}
\newcommand{\eea}{\end{eqnarray}}
\newcommand{\bef}{\begin{figure}}
\newcommand{\eef}{\end{figure}}

\newcommand{\rpp}{\rho \pi \pi}

\newcommand{\eps}{\epsilon}

\newcommand{\lgl}{\langle}
\newcommand{\rgl}{\rangle}

\begin{document}
\title{\large\bf Study of Spectral Modification of Hadrons through 
Electromagnetic Probes\footnote{Presented at DAE Symposium on Nuclear
Physics, Kolkata, India, 26-30 Dec. 2001}}
\author{\bf Sourav Sarkar}

\address{Variable Energy Cyclotron Centre \\1/AF Bidhannagar, Kolkata-700 064}

\maketitle
\abstracts{The medium modification of vector meson properties and its effects
on thermal photon and dilepton emission rates from hot/dense hadronic
matter are discussed. The role of spectral changes of hadrons in explaining
the WA98 photon data and the CERES/NA45 dilepton data with similar initial 
conditions is demonstrated.}

\vskip 0.2in
\section{Introduction}

A fundamental property of QCD with two light quark flavours
is the approximate chiral
symmetry which plays a very important
role in the understanding of low energy hadron physics.
Since this symmetry is not manifest in the observed particle
degeneracies it is believed that chiral symmetry is spontaneously
broken by the vacuum. Under conditions of high
temperature or density one expects that the QCD vacuum will undergo
substantial modifications leading to the restoration of
chiral symmetry.  The chiral condensate which has a non-trivial 
value in the broken phase acts as the order parameter of this transition.
Though hadron masses can be expressed in terms of various condensates
through the QCD sum rules for example, there are uncertainties connected with
the approximations used in their evaluation. The 
change in the condensate in hot/dense environment is expected to show up  
in the spectral function of
the vector mesons. Concomitant to these phenomena, quarks and gluons, 
which are normally bound within hadrons are expected to undergo
a deconfinement transition at sufficiently high temperature or
baryon density leading to the formation of quark gluon plasma (QGP)~\cite{qm99}.

Relativistic collisions of heavy ions provide a rich arena for the 
study of strongly interacting matter under extreme conditions
in the laboratory.
Data from the Super Proton
Synchrotron (SPS) have already been analysed and the first results from
the Relativistic Heavy Ion Collider (RHIC) have started to appear.
Electromagnetic probes, {\it viz.}, photons
and dileptons have long been recognized as the most direct probes
of the collision~\cite{emprobe}.
Owing to the nature of their interaction they undergo minimal scatterings
and are by far the best markers of the entire space-time evolution
of the collision.
Both photons and dileptons
couple to hadrons through spin one ~($J^{P}=1^-$)
mesons. The dilepton spectra in particular, exhibits a resonant structure
which, in the low mass regime includes the $\rho$ and
the $\omega$ mesons.
It has been emphasized that the properties of the vector mesons
undergo nontrivial medium modifications in a hot and/or dense medium
such as likely to be produced in relativistic nuclear collisions.
Consequently the spectral modifications of these vector mesons
would be clearly revealed in the invariant mass spectra of the dileptons
through the shift of the resonant peaks.
\par
In the following we will discuss the role of medium 
modification of hadronic properties in explaining the photon and dilepton
spectra from heavy ion collisions at the CERN SPS.
Let us first consider the data
obtained from Pb-Pb collisions at 158 A GeV
reported by the WA98 Collaboration~\cite{wa98}
which shows the transverse momentum spectra of excess photons above those
coming from the decay of $\pi^0$ and $\eta$ mesons. 
These excess photons are believed to arise due to (a) the hard collisions
of partons in the colliding nuclei and (b) a thermal medium - QGP or
hot/dense hadronic matter. We have seen~\cite{rapid}
that the above spectra can be
explained using a reasonable set of parameters if the medium effects of
the vector meson properties are consistently incorporated in the 
invariant rates as well as in the equation of state.
It is worth recalling that the single photon data for S+Au collisions could
also be explained in a similar scenario~\cite{sarkar}.
However, other attempts to explain the photon spectra without
invoking medium effects exist in the literature~\cite{cywong,kg,dyp,dks,pasi}.
\par
Our aim now is to check whether the same scenario can explain the dilepton
data obtained in Pb+Au collisions at SPS by the CERES/NA45 
Collaboration~\cite{ceres}.
The possible sources in the low mass region are the dileptons coming
from the decays of hadrons at freeze-out and from the in-medium
propagation and decay of vector mesons. The data shows a significant
enhancement in the mass region 0.3 to 0.6 GeV which can 
be explained~\cite{lowmass} by a substantial negative shift of the mass 
of vector mesons (the $\rho$
meson in particular) in the thermal medium.
A large broadening of the $\rho$ meson spectral function due to scattering off
baryons can also explain this enhancement~\cite{rapp}.
\par
In the following we will see that
both the WA98 photon spectra  as well as the CERES/NA45 dilepton spectra
for higher multiplicities can be explained by either of the two scenarios of 
relativistic nuclear collision~\cite{red,rapid}:
(a) A\,+A\,$\ra$QGP$\ra$Mixed Phase$\ra$Hadronic Phase
or (b) A\,+\,A\,$\ra$Hadronic Matter;
with downward shift of vector meson masses
and initial temperature $\sim 200$ MeV. The
photon yield is seen to be insensitive to the broadening
of vector mesons~\cite{annals},
although the CERES dilepton data admits such a scenario
as mentioned above. The effects of the thermal shift of the hadronic
spectral functions on both photon and dilepton emission have
been considered in Ref.~\cite{annals}
for an exhaustive set of models.  An appreciable
change in the space-time integrated yield of electromagnetic
probes was observed for universal scaling
and Quantum Hadrodynamic (QHD) model and we will consider only these
in our discussion.
\par This report is organized as follows. In the following section we will
briefly outline the phenomenology of medium effects on the vector mesons
in the thermal environment which we have considered. In section~3 we will
consider the static photon and dilepton rates due to different processes. 
Then in
section~4 we will describe the dynamics of space-time evolution followed by the
results of our calculation in section~5. We will conclude with a
summary and discussions in section 6.

\section{Medium Effects}

A substantial amount of literature has been devoted  to the
issue of temperature and/or density dependence of hadrons within
various models~\cite{annals,thpr,brpr,rapp,rdp}. 
In this work we will consider medium modifications of vector mesons
in two different scenarios :
the universal scaling hypothesis
and the Quantum Hadrodynamic (QHD) model.

In the scaling hypothesis, the  parametrization of in-medium
quantities (denoted by $*$) at finite temperature, $T$ and
baryon density~\cite{geb}, $n_B$ is
\be
{m_{V}^* \over m_{V}}  =
{f_{V}^* \over f_{V}} =
{\omega_{0}^* \over \omega_{0}}  =\left(1-0.2\frac{n_B}{n_B^0}\right)
 \left( 1 - {T^2 \over T_c^2} \right) ^{\lambda},
\label{anst}
\ee
where $V$ stands for vector mesons, $f_V$ is
the coupling between the electromagnetic current and the vector
meson field, $\omega_0$ is the continuum threshold,
$T_c$ is the critical temperature and $n_B^0$ is the
baryon density of normal nuclear matter.
Mass of the nucleon also varies with temperature
according to Eq.~(\ref{anst})
(pseudo scalar masses remain unchanged).
The particular exponent $\lambda=1/2$ (1/6)
is known as Nambu (Brown-Rho) scaling~\cite{brpr}. We will use $\lambda=1/2$
in our calculations.
Note that there is no definite reason to believe that all the in-medium
dynamical quantities are dictated by a single exponent $\lambda$.
This is the simplest possible ansatz (see~\cite{annals} for a discussion).
The effective mass of $a_1$
is estimated by using Weinberg's sum rules~\cite{weinberg}.

In the Quantum Hadrodynamic model~\cite{vol16,chin} of nuclear matter
the vector meson properties are modified due to coupling with
nucleonic excitations.
The nucleons interact through the exchange
of scalar $\sigma$ and the vector $\omega$
mesons and their mass is modified due to the scalar condensate.
This is evaluated in the Relativistic Hartree Approximation (RHA).
Coupling with these modified nucleonic excitations induce
changes in the $\rho$ and $\omega$ meson masses.
This modification is contained in the meson self energy
which appears in the Dyson-Schwinger equation for the effective propagator
in the medium. The interaction vertices are provided by the Lagrangian
\be
{\cal L}^{\s int}_{VNN} = g_{VNN}\,\left({\bar N}\gamma_{\mu}
\tau^a N{V}_{a}^{\mu} - \frac{\kappa_V}{2M_N}{\bar N}
\sigma_{\mu \nu}\tau^a N\partial^{\nu}V_{a}^{\mu}\right),
\label{lagVNN}
\ee
where $V_a^{\mu} = \{\omega^{\mu},{\vec {\rho}}^{\mu}\}$,
$M_N$ is the free nucleon mass, $N$ is the nucleon field
and $\tau_a=\{1,{\vec {\tau}}\}$, $\vec\tau$ being the Pauli matrices.
The real part of the vector meson self-energy due to $N\bar N$ polarization
is responsible for the mass shift (see~\cite{npa1,npa2,csong,kuwabara}).
The $\rho-\pi$ interaction is given by
\be
{\cal L}^{int}_{\rpp}=
-g_{\rpp}\vec{\rho_\mu}\cdot (\vec{\pi}\times\partial^\mu\vec{\pi})
+\frac{1}{2}g_{\rpp}^2\,
(\vec\pi\times\vec\rho_\mu)\cdot(\vec\pi\times\vec\rho^\mu).
\label{lagrpp}
\ee
This makes the dominant contribution to the imaginary part 
of the $\rho$ self-energy 
though the pole shift (real part) is negligibly small.
We have evaluated the spectral function of $\rho$ due to
its coupling with the pion and nucleons
(Eqs.~(\ref{lagVNN}) and (\ref{lagrpp}))
in the thermal bath for finite three momentum at
non-zero  temperature and  density.

\section{Thermal Photon and Dilepton Emission Rates}

In this section, we briefly recapitulate the main
equations relevant for evaluating photon and dilepton emission
from a thermal source.
The emission rate of real photons with four momentum $p_\mu\,=(E,\vec p)$
can be expressed in terms of the imaginary part of
the retarded electromagnetic current correlation function
as~\cite{emprobe}
\be
E\frac{dR}{d^3p}=-\frac{1}{(2\pi)^3}
\,g^{\mn}\,{\s Im}W^R_{\mn}(p)
\,f_{BE}(E,T)\,,
\label{drpho}
\ee
where
\be
W^R_{\mu\nu}(p)\equiv i\int\,d^4x\,\,e^{ip\cdot x}\,\theta(x_0)\,\langle[J^{em}_\mu(x),
J^{em}_\nu(0)]\rangle_\beta\,,
\ee
$J^{em}_\mu$ being the electromagnetic current of quarks or hadrons.
All the information about the thermal
medium which produces the photons or dileptons resides in the current
correlator $W^{\mn}$.
The lowest order processes contributing to thermal
photon emission from quark
gluon plasma are the QCD Compton and annihilation
processes~\cite{kapusta}. It has been
shown recently~\cite{aurenche} that the two-loop contribution leading to
bremsstrahlung and $q\bar q$ annihilation with scattering is of
the same order as the lowest order processes. The total rate of emission
per unit four-volume at temperature $T$ is given by
\bea
E\frac{dR}{d^3p}&=&\frac{5}{9}\,\frac{\alpha\alpha_s}{2\pi^2}\,\,
\exp(-E/T)
\,\left[\ln\left(\frac{2.912\,E}{g^2\,T}\right)\right.
\nonumber\\
&&\left.+4\frac{(J_T-J_L)}{\pi^3}
\{\ln2+\frac{E}{3T}\}\right]
\label{edr}
\eea
where $J_T\simeq4.45$ and $J_L\simeq-4.26$.

Considering the hadronic matter to be made up of $\pi$, $\rho$, $\omega$, 
$\eta$ and $a_1$ mesons, the imaginary part of $W^{\mn}$ gives the amplitudes
of a number of hadronic processes. We have considered the reactions
$\pi\,\rho\,\ra\, \pi\,\gamma$,
$\pi\,\pi\,\ra\, \rho\,\gamma$, $\pi\,\pi\,\ra\, \eta\,\gamma$,
$\pi\,\eta\,\ra\, \pi\,\gamma$ and the decays $\rho\,\ra\,\pi\,\pi\,\gamma$
and $\omega\,\ra\,\pi\,\gamma$ with dressed vector propagators
to estimate photon emission from hadronic matter.

To obtain the dilepton emission rate, the additional part to be included in 
Eq.~(\ref{drpho}) corresponds to the free propagation of the
virtual photon and its subsequent decay to lepton pairs. This gives,
\be
\frac{dR}{d^4p}=-\frac{\alpha}{12\pi^4\,p^2}
\,g^{\mn}\,{\s Im}W^R_{\mn}(p)
\left(1+\frac{2m_l^2}{p^2}\right)\,\sqrt{1-\frac{4m_l^2}{p^2}}
\,f_{BE}(E,T)
\label{drlep}
\ee
where $m_l$ is the lepton mass.

In the QGP, the dominant channel for dilepton productions is
quark-antiquark annihilation.
The rate for this process is obtained from the lowest order
diagram contributing to the current correlator $W^{\mn}$ and
is obtained as,
\be
\frac{dR}{d^4p}=\sum_f e_f^2\,\frac{\alpha^2}{4\pi^4}
\left(1+\frac{2m_l^2}{p^2}\right)\,\sqrt{1-\frac{4m_l^2}{p^2}}
\,f_{BE}(E,T)
\label{qqbarlep}
\ee

Now let us consider dilepton production in the hadronic medium.
In order to obtain the rate of dilepton emission from hadronic matter
($\rho/\omega\ra l^+l^-$) from Eq.~(\ref{drlep})
the electromagnetic current correlator is expressed in terms of
the effective propagator of the vector particle in the thermal medium
using vector meson dominance (VMD) so as to obtain
\bea
{\s Im}W_{\mu\mu}^R&=&\frac{e^2 m_V^4}{g_V^2}
\left[\frac{2{\s Im}\Pi_V^T}{(p^2-m_V^2+{\s Re}\Pi_V^T)^2
+({\s Im}\Pi_V^T)^2}\right]\nonumber\\
&&~~~~~~~~~~~~~~~\left.+\frac{{\s Im}\Pi_V^L}{(p^2-m_V^2+{\s Re}\Pi_V^L)^2
+({\s Im}\Pi_V^L)^2}\right].
\eea
where ${\s{Im}}\Pi_V^{T(L)}$ is the transverse(longitudinal) part of
the retarded self-energy of vector mesons arising out of interaction with
excitations in the medium. The dilepton rate from pion annihilation for
example, can be obtained from $\rho$ self-energy due to $\pi-\pi$ polarisation 
in the medium.
In the approximation $\Pi^T_{\rho }=\Pi^L_{\rho }=\Pi_\rho$ we have
\be
{\s Im}W_{\mu\mu}^R=\frac{3e^2 m_\rho^{\ast 4}}{g_{\rpp}^2}
\left[\frac{{\s Im}\Pi_{\rho}}{(M^2-m_\rho^{\ast 2})^2
+({\s Im}\Pi_{\rho})^2}\right],
\label{walcor}
\ee
where $m_\rho^{\ast 2}
=m_\rho^2-{\s Re}\Pi_\rho$ and $p^2=M^2$, the invariant mass of the lepton
pair. This is proportional to the familiar electromagnetic form factor
of the pion.

\section{Space-Time Evolution}

The observed spectrum originating from an expanding
hadronic matter is obtained by convoluting the static
rates given above with the expansion dynamics. 
This is done using (3+1) dimensional relativistic hydrodynamics
assuming boost invariance in the longitudinal direction 
and cylindrical symmetry in the transverse plane.
The initial energy density and transverse velocity profiles are taken as
\be
\eps(\tau_i,r)=\frac{\eps_0}{e^{(r-R_A)/\delta}+1};~~~~~
v_r=v_0\left(\frac{r}{R_A}\right)^\alpha
\label{velprofile}
\ee
with $\alpha=1$.
The other essential input one requires is
the equation of state (EOS) which provides the cooling law.
For the QGP sector we use the bag model equation of state with
two flavour degrees of freedom. 
The hadronic phase is taken to consist of $\pi$, $\rho$, $\omega$,
$\eta$ and $a_1$
mesons and nucleons.
The medium effects enter through the effective masses in
the expressions for energy density and pressure.
The entropy density is then parametrized as,
\be
s_H=\frac{\epsilon_H+P_H}{T}\,\equiv\,4a_{\s{eff}}(T)\,T^3
= 4\frac{\pi^2}{90} g_{\s{eff}}(m^\ast,T)T^3
\label{entro}
\ee
where  $g_{\s{eff}}$ is the effective statistical degeneracy.
Thus, we can visualize the finite mass of the hadrons
having an effective degeneracy $g_{\s{eff}}(m^\ast,T)$.
The velocity of sound which plays a significant role in the expansion,
also becomes a function of $T$ and differs substantially from its value
corresponding to an ideal gas ($1/\sqrt 3$). 
The initial temperature of the system is obtained by solving
the equation
\be
T_i^3=\frac{3.6}{\pi R_A^2\,\, 4a_{\s{eff}}(T_i)\,
\tau_i}\,\frac{dN_\pi}{dy}
\label{dndy}
\ee
where $dN_\pi/dy$ is the total pion multiplicity
($\sim 1.5\times$ charge multiplicity), $R_A$ is the radius
of the system, $\tau_i$ is the initial thermalization time and
$a_{\s{eff}}=({\pi^2}/{90})\,g_{\s{eff}}(T_i)$.
When the system is produced in the QGP phase, $g_{\s{eff}}$ 
is replaced by $g_{QGP}$ which
for two quark flavours is 37. If the quark and gluon
masses are non-zero in the thermal bath then the effective
degeneracy of the QGP, $g_{\s{QGP}}^{\s{eff}}$, defined by Eq.~(\ref{entro})
can be lower than 37, resulting in a higher value of $T_i$ for
a given multiplicity according to Eq.~(\ref{dndy}).
Note that the change in the expansion dynamics
as well as the value of the initial temperature due
to medium effects relevant for a hot hadronic gas
also enters through the effective statistical degeneracy.
The freeze-out temperature $T_F$ is taken as 120 MeV.

\section{Results}
%%%%%%%%%%%%%% Fig. 1 %%%%%%%%%%%%%%%%%%%%%%%%%%%%%
\bef
\centerline{\psfig{figure=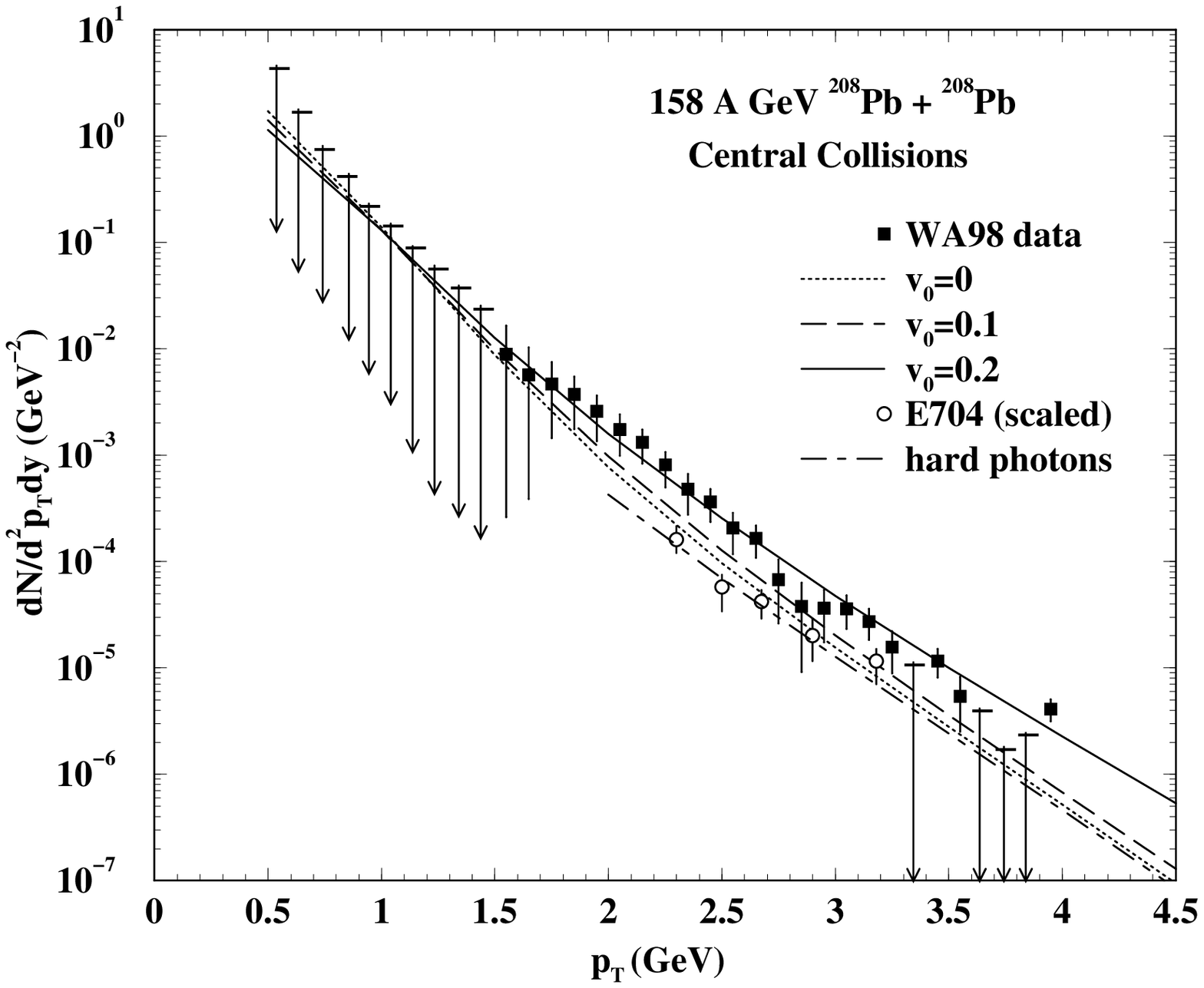,height=5cm,width=7cm}}
\caption{Total photon yield in Pb + Pb collisions
at 158 A GeV at CERN-SPS. 
The system is formed in the QGP phase with initial temperature
$T_i=190$ MeV.
}
\label{fig2}
\eef
%%%%%%%%%% End of Fig. 1 %%%%%%%%%%%%%%%%%%%%%%%%%%%%%%%%

In Fig.~\ref{fig2},
the photon yield with QGP in the initial state
is shown for three different values of the initial transverse velocity.
The yield from hadronic matter during the mixed and
hadronic phases are calculated with medium effects in the scaling scenario.
All the three curves represent the sum of the thermal and
the prompt photon contribution which includes possible finite $k_T$
effects of the parton distributions. The later, shown separately by
the dot-dashed line also explains the scaled $pp$ data from E704
experiment~\cite{e704}.
We observe that the photon spectra for the initial velocity
profile given by Eq.~(\ref{velprofile}) with  $v_0=0.2$ explains
the WA98 data reasonably well. 
It is found that a substantial fraction of the photons come from mixed and
hadronic phase. The contribution from the QGP phase
is small because of the small life time of the
QGP ($\sim 1$ fm/c).
 
%%%%%%%%%%%%%% Fig. 2 %%%%%%%%%%%%%%%%%%%%%%%%%%%%%
\bef
\centerline{\psfig{figure=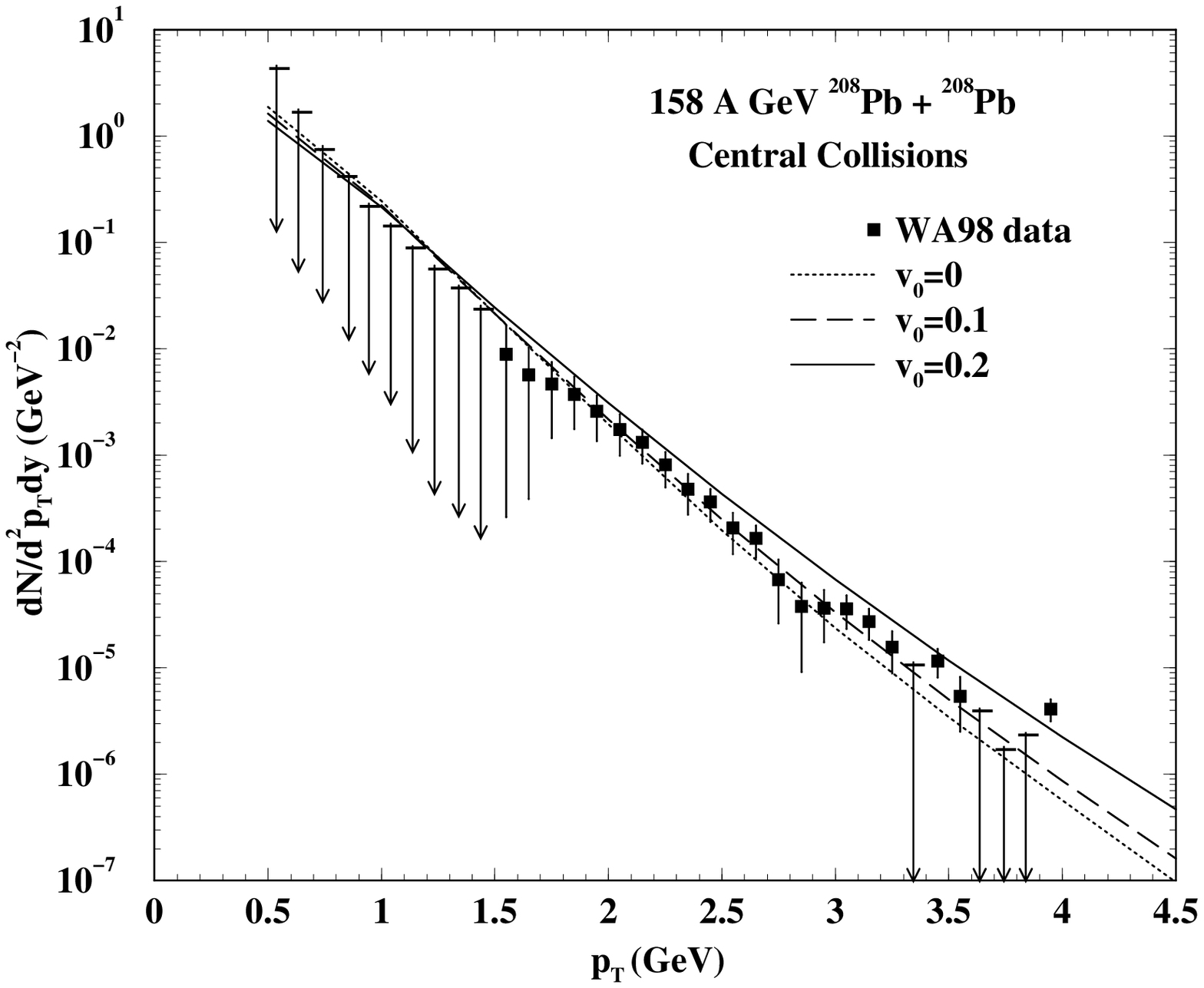,height=5cm,width=7cm}}
\caption{
Total photon yield in Pb + Pb collisions
at 158 A GeV at CERN SPS. 
The system is formed in the hadronic phase with
the hadronic masses approaching zero at initial temperature
$T_i=200$ MeV.
}
\label{fig3}
\eef
%%%%%%%%%% End of Fig. 2 %%%%%%%%%%%%%%%%%%%%%%%%%%%%%%%%

 To check whether 
 the  existence of the QGP phase essential to reproduce the WA98 data,
 we have considered two possibilities: 
(a) pure hadronic model without medium-modifications,
and (b) pure  hadronic model  with scaling hypothesis according to
Eq.(\ref{anst}).  In the former case,
$T_i$ is found to be $\sim 250 $ MeV for
$\tau_i=1$ fm/c and $dN/dy=700$,
 which appears to be too high for the hadrons to survive. Therefore
  this possibility should be excluded.
  On the other hand,
 the second case with an assumption of $T_i = T_c$ (which is
  just for simplicity) leads to
  $T_i\sim 200$ MeV, at $\tau_i= 1$ fm/c, which is not unrealistic.
In this case, the hadronic
system expands and cools and ultimately
freezes out at $T_f$=120 MeV.
The results 
for three values of the initial radial velocity including the prompt photon
contribution are shown in Fig.~\ref{fig3}.
The experimental data are well reproduced for vanishing
initial transverse velocity also.
This indicates that a simple hadronic model is inadequate.
 Either substantial medium modifications
of hadrons or the formation of QGP in the initial stages is necessary to
reproduce the data. It is rather difficult to distinguish
between the two at present.

We will now compare the results of our calculation with the dilepton 
spectra obtained by the CERES/NA45 Collaboration.
We have considered~\cite{lowmass} dileptons with transverse
momentum $p_T$ above 200 MeV/c
and an opening angle $\Theta_{ee}>$ 35 mrad.
These are kinematical cuts relevant for the CERES detector
and are incorporated as described in ~\cite{solfrank}.
In all the figures, the quantity $\lgl N_{ch}\rgl$
indicates the average number
of charged particles per unit rapidity interval in the pseudorapidity
range $2.1<\eta<2.65$. In all the cases background due to hadron
decays are added to the thermal yields.

%%%%%%%%%%%%%% Fig. 4 %%%%%%%%%%%%%%%%%%%%%%%%%%%%%
\bef
\centerline{\psfig{figure=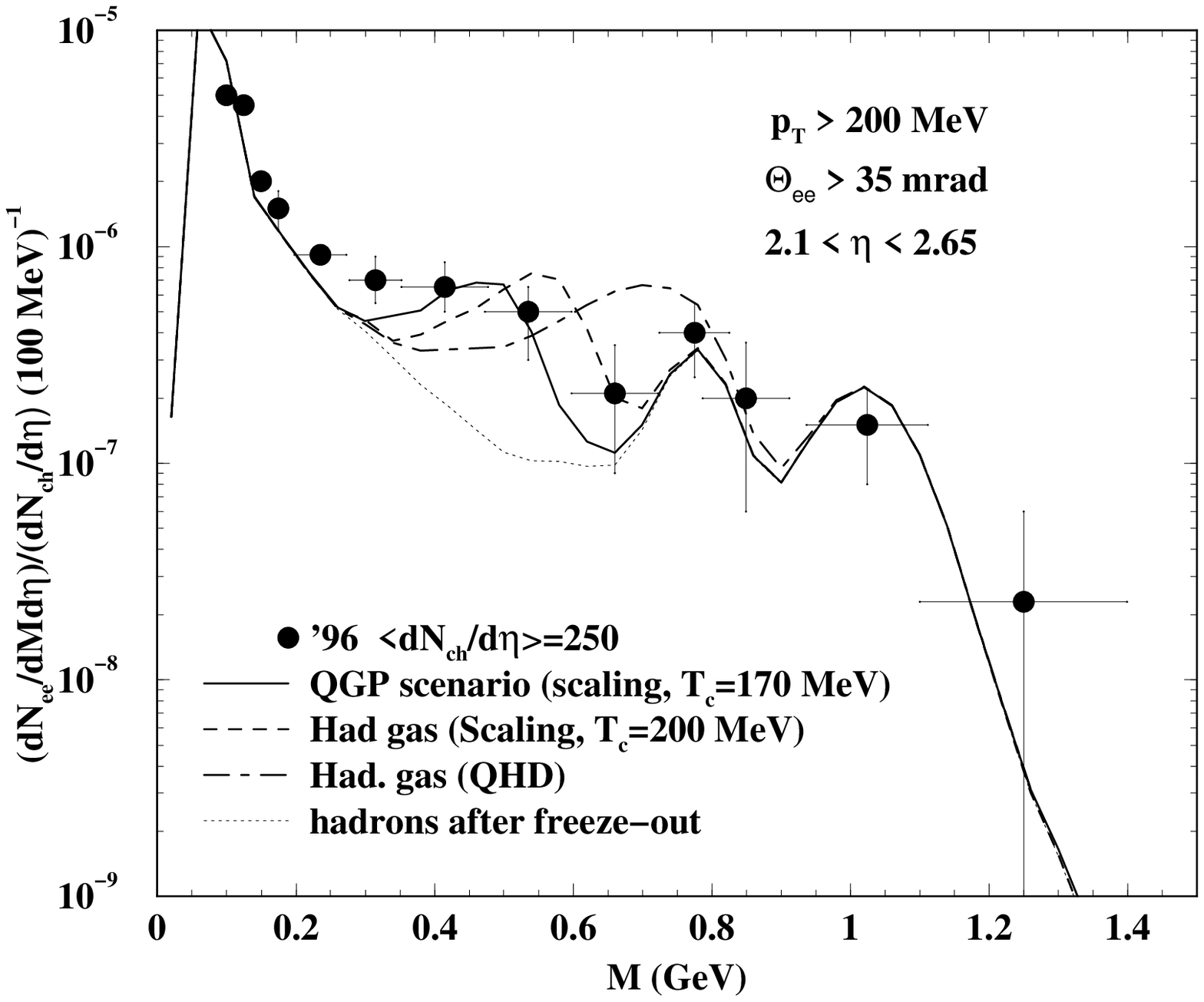,height=5cm,width=7cm}}
\caption{Dilepton spectra for $\lgl N_{ch}\rgl$=250 for
different initial states and mass variation scenarios.
}
\label{f250}
\eef
%%%%%%%%%% End of Fig. 4 %%%%%%%%%%%%%%%%%%%%%%%%%%%%%%%%

We display the results for $dN_{ch}/d\eta=250$ in Fig.~\ref{f250}.
In this case apart from the hadronic gas
scenario we have also considered a deconfined initial state with
 $T_i=180$ MeV
which
evolves into a hadronic gas via a mixed phase.
The observed enhancement of the dilepton yield
around $M\sim 0.3 - 0.6 $ GeV can be reproduced
with the QGP initial state, once the variation
of vector meson masses in the mixed and the hadronic phases
are taken into account (solid line).
The data is also reproduced
by a hadronic initial state with $T_i=190$ MeV 
in the universal mass variation scenario (dashed line).
The $\rho$-peak in the dilepton spectra shifts towards
lower $M$ for universal scaling compared to QHD model, indicating
a strong medium effect in the former case. 

The dilepton spectra
for QGP and hadronic initial states for $dN_{ch}/d\eta=350$ is shown
in Fig.~\ref{f350}.
Results with
hadronic initial state and universal scaling of
hadronic masses with temperature describes the data
reasonably well.
We see that with the temperature dependent mass from QHD
model the low mass enhancement of the experimental
yield cannot be reproduced.
A good description of the data
can be obtained by taking $T_i=200$ MeV
with QGP initial state for $T_c=190$ MeV.
We also show the results due to large broadening
of the $\rho$ spectral function in the medium. The broadening
of $\rho$ is modelled by assuming the temperature dependent
width as : $\Gamma_\rho(T)=\Gamma_\rho(0)/(1-T^2/T_c^2)$.

%%%%%%%%%%%%%% Fig. 5 %%%%%%%%%%%%%%%%%%%%%%%%%%%%%
\bef
\centerline{\psfig{figure=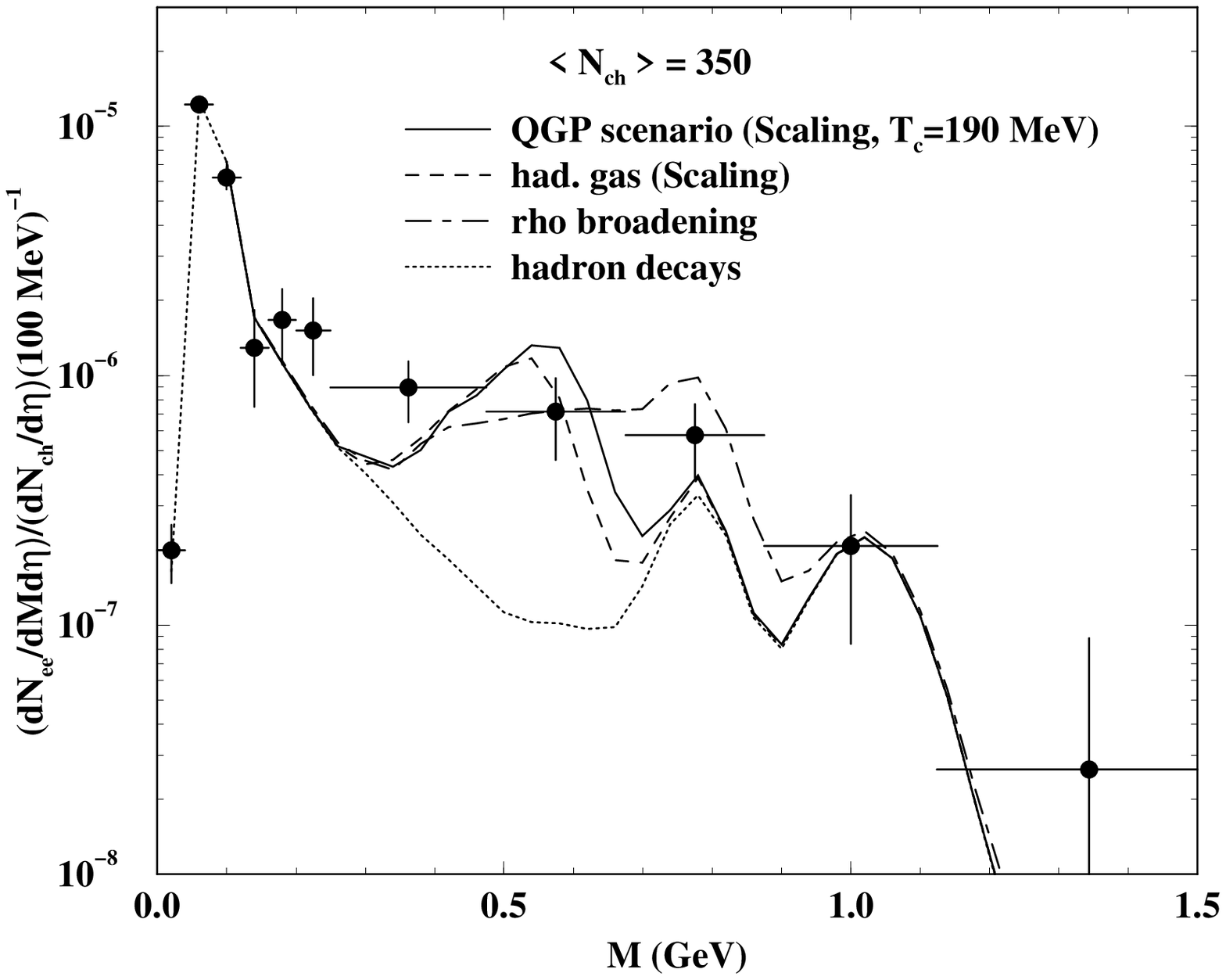,height=5cm,width=7cm}}
\caption{Dilepton spectra for $\lgl N_{ch}\rgl$=350 for
different initial states and mass variation scenarios.
}
\label{f350}
\eef
%%%%%%%%%% End of Fig. 5 %%%%%%%%%%%%%%%%%%%%%%%%%%%%%%%%

\section{Summary and Discussions}

We have evaluated the high energy photon yield
in Pb + Pb collisions at
CERN SPS energies with two different initial conditions.
In the first scenario, we start with the assumption of the formation
of a QGP phase at an initial temperature of
$\sim 200$ MeV and then the system continues through
mixed phase and hadronic phase before freeze-out. In the second
scenario, we assume a chirally symmetric phase where the hadronic masses
approach zero at a temperature $\sim 200$ MeV and then the system
evolves towards freeze-out.
The effects of the variation of hadronic masses
on the photon yield have been considered both in the cross section
as well as in the equation of state.
The photon spectra reported by the WA98 collaboration are well
reproduced in both cases.
We have also studied the dilepton yield
measured by CERES experiment
for two values of the charge multiplicity
in Pb + Au interactions. It is
observed that 
the data can be described by both QGP and hadronic
initial states with an initial temperature $\sim$ 200 MeV. 
We have assumed $\tau_i=1$ fm/c at SPS energies,
which may be considered as the
lower limit of this quantity,  because
the transit time (the time taken by the nuclei to pass
through each other in the CM system) is $\sim$ 1 fm/c at SPS
energies and
the thermal system is assumed to be formed after this time
has elapsed.
There are also uncertainties in the value of $T_c$~\cite{lattice},
a value of $T_c\sim 200$ MeV may be considered as an
upper limit.  Moreover, the photon
emission rate from QGP given by Eq.~(\ref{edr}), evaluated in
Refs.~\cite{kapusta,aurenche}
by resumming the hard thermal loops is strictly valid for $g<<1$
whereas the value of $g$ is $\sim 2$
at $T\sim 200 $ MeV. At present it is not clear whether the rate
in Eq.~(\ref{edr}) is valid for such a large value of $g$.
\vskip 0.1in
\noindent{{\bf Acknowledgement}:
The work reported here was done in collaboration with J. Alam, P. Roy, 
T. Hatsuda, T. K. Nayak and B. Sinha at various stages.

\end{document}